\documentclass[aps,prl,epsfigure,showpacs,twocolumn]{revtex4}
\usepackage{dcolumn}
\usepackage{bm}
\usepackage{graphicx}
\usepackage{amsmath}
\usepackage{latexsym}
\usepackage{amsfonts}
\usepackage{amssymb}
\usepackage{array}
\usepackage{times}

\newcommand{\ket}[1]{\left\vert#1\right\rangle}

\newcommand{\bra}[1]{\left\langle#1\right\vert}

\begin{document}
\title{Solitons in interacting Dicke models of coupled cavities with two-level systems}
%\title{Solitonic behavior in coupled multi atom-cavity systems}
\author{M. Paternostro$^{1}$, G. S. Agarwal$^2$, and M. S. Kim$^1$}
\affiliation{$^1$School of Mathematics and Physics, Queen's University, Belfast BT7
  1NN, United Kingdom\\
$^2$Department of Physics, Oklahoma State University, Stillwater, Oklahoma 74078, USA}
\date{\today}
\begin{abstract}
We consider an array of coupled optical cavities, each containing a multi-atom ensemble. We show that the nonlinearity inherent in the cooperative dynamics of the atoms in each ensemble coupled to the respective cavity field allows for the formation of solitary waves. Such a prediction can be tested in state-of-the-art semiconducting photonic-crystal microcavities with embedded impurities.
\end{abstract}
 
\pacs{03.75.Lm, 42.65.-k, 42.50.Pq, 42.65.Tg}

\maketitle

Very recenty, a considerable amount of interest has been directed towards registers of coupled atom-cavity systems as potential candidates for the observation of cooperative phenomena in strongly correlated many-body systems~\cite{generale}. Superfluid-to-Mott-insulator quantum phase transitions and glassy polaritonic phases have been theoretically predicted in arrays of mutually coupled qubit-resonator devices. At the basis of these interesting effects is the nonlinear character of the so-called polariton, a combined state of an atom and a photon. Moreover, the flexibility of these systems, which makes them exploitable quantum simulators, can be used in order to realize effective multiple-spin dynamics useful for the purposes of quantum information processing~\cite{simulator}.

However, the possibilities offered by arrays of cavities interacting with atoms
%which have the potential to narrow the gap between prediction and exploitation of non-linear phenomena for quantum information processing taks, 
are enormous and intriguing.
%is certainly not exhausted by the investigation of polaritonic systems. 
For instance, photonic Mott insulators can be achieved from atom-mediated nonlinear photon-photon interactions~\cite{hartmannnonlinear}. In general, the combination of strong {\it intracavity} atom-photon interactions and {\it intercavity}  tunneling is greatly suitable for nonlinear-optics purposes. 

Here we further explore these possibilities and report for the first time the formation of solitary waves (or {\it solitons})~\cite{agrawal,christo} in a periodic array of Dicke systems~\cite{books}. It is believed that solitons, which are localized nonlinear waves characterized by striking stability properties with respect to dynamical perturbations (such as collisions with other solitons) represent, together with squeezing, one of the most remarkable manifestations in nonlinear optics~\cite{phystoday}. The situation we consider is different from a classical array of media with Kerr nonlinearity as, technically, we do not expand the atomic polarization as a power-series of the field or assume fast response of the atomic system relative to the field. Differently, we treat the dynamics of both cavity fields and atomic ensembles on equal footing, our source of nonlinearity being intrinsic to each ensemble of atoms. The identification of solitonic behavior in a system of current theoretical as well as experimental interest is an important step forward in the grounding of coupled-cavity registers as reliable and exploitable quantum simulators. We show that, by embedding a small number of two-level systems within a cavity constituting one site of an array, it is possible to achieve a degree of nonlinearity which, combined with the periodicity of the cavity arrangement, allows for the formation of solitons. Our results are derived by means of a simple technique permitting the consideration of any desired/appropriate order of nonlinearity.
% in the coupling between each cavity field and the corresponding atomic ensemble. 
Furthermore, although for simplicity we consider a linear array of coupled cavities, the extension to bidimensional configurations is certainly possible. We identify in semiconducting photonic crystal microcavities (doped with multiple impurities) a foreseeable potential setup for our proposal~\cite{pc}. By requiring a small number of impurities per cavity, our proposal demonstrates the interest of a working-point which is intermediate with respect to 
%the extremes represented by 
the simulation of the Bose-Hubbard model (expected when each cavity is off-resonantly interacting with a large number of two-level system) and the fully microscopic polaritonic dynamics~\cite{generale}.

\begin{figure}[t]
%  {\bf (a)}\hskip3.5cm{\bf (b)}
\centerline{\includegraphics[width=0.37\textwidth]{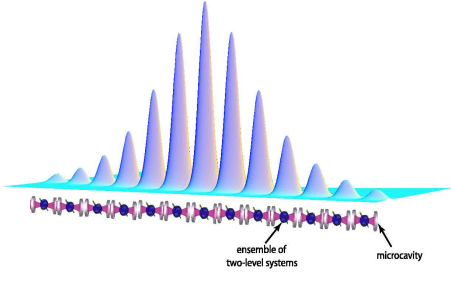}}
\caption{Simplified schematics of the setup and representation of a possible descrete solitonic solution of the dynamics associated with the cavity fields (see the body of the paper). Each resonator ideally represents a defect microcavity in a photonic crystal, the corresponding ensemble of two-level atoms being embodied by impurities.}
\label{fig:fig0}
\end{figure}

{\it The physical model}.- In order to clearly elucidate the salient features of the mechanism under consideration, we address the simple case of a linear array of mutually coupled microcavities. For the sake of simplicity, we describe the cavities as $M$ equally spaced and independent optical resonators, each occupying a site of the array. 
%An experimentally feasible physical setup is discussed later on. 
Each cavity contains $N$ identical two-level systems whose ground (excited) state we denote as $\ket{g}$ ($\ket{e})$. The specific nature of the two-level systems is immaterial for the following duscussion. Therefore, for easiness of language, we refer to them as {\it atoms} until we address the physical setup we suggest. The cavities are sufficiently close to each other to allow for their mutual coupling through evanescent-photon hopping (see Fig.~\ref{fig:fig0} for a sketch). The cavity-coupling drops exponentially with the distance so that we consider only nearest-neighbor interactions. 
%However, it is important to stress that the geometry of the array can be taken to be bidimensional without loss of generality of the following discussions. 

Each atom interacts with the respective cavity via the electric-dipole coupling. As the volume of each cavity can be made very small ($\sim\lambda_c^3$ with $\lambda_c$ the wavelength of the cavity field), the coupling within each cavity can be large
%. In what follows, we assume the experimentally reasonable condition of negligible spontaneous emission from the two-level atoms
~\cite{yama}. 
The free Hamiltonian of the whole system is $\hat{\cal H}_{0}=\sum^M_{j=1}(\omega_{c}\hat{a}^\dag_{j}\hat{a}_j+\omega_{eg}\sum^N_{k=1}\hat\sigma^z_{kj})$, where $\omega_{c}$ is the frequency of the cavity fields, $\omega_{eg}$ is the transition frequency of the atoms, $\hat{a}^{\dag}_{j}$ is the creation operator of mode $j$ and $\hat{\sigma}^z_{kj}$ is the $z$-Pauli matrix for the $k$-th atom interacting with the $j$-th cavity of the array [we use units such that $\hbar=1$ throughout the paper]. In the resonant condition $\omega_{c}\simeq{\omega}_{eg}$ and in the interaction picture with respect to $\hat{\cal H}_0$, the coupling Hamiltonian reads
\begin{equation}
\label{hamiltonian}
%\hat{\cal H}=-\sum^M_{j=1}J_j(\hat{a}_{j}\hat{a}^\dag_{j+1}+\hat{a}^\dag_{j}\hat{a}_{j+1})+\sum^M_{j=1}\Omega_j(\hat{a}_{j}\hat{S}^+_{j}+a^\dag_j\hat{S}^-_j).
\hat{\cal H}=-\sum^M_{j=1}J_j\hat{a}_{j}\hat{a}^\dag_{j+1}+\sum^M_{j=1}\Omega_ja^\dag_j\hat{S}^-_j+h.c.
\end{equation}
Here, we have introduced the {\it Dicke lowering operator} of the atoms within a given cavity $\hat{S}^-_{j}=\sum^N_{k=1}\hat{\sigma}^-_{kj}$ (with $\hat{\sigma}^{-}_{kj}=\ket{g}_{kj}\!\bra{e}$ and $\hat{S}^+_j=\hat{S}^{-\dag}_{j}$), the photon-hopping strength $J$ and the coupling rate $\Omega$ between an ensemble and its respective cavity. To fix the ideas, these parameters are taken to be real and homogeneous across the array so that $J_j=J$ and $\Omega_j=\Omega,~\forall{j}$, although the generalization to a {\it pattern} of couplings is quite straightforward. The first term (and its $h.c.$) in $\hat{\cal H}$ describes the tunneling-like coupling between the cavities while the second term accounts for the {\it in situ} resonant exchange of excitations between the $N$ elements of an ensemble and the corresponding cavity field (energy non-conserving terms have been discarded by invoking the rotating wave approximation, valid for $\Omega\ll{\omega_c}$). 

The extended character of the Hilbert space of the overall system makes the goal of tackling the dynamics described by $\hat{\cal H}$ formidable. However, we can exploit the collective behavior of each 
%e two-level atoms forming a given 
ensemble by utilizing the Holstein-Primakoff (HP) transformation, which maps a physical collective spin into an effective bosonic degree of freedom~\cite{HP}. Assuming $N\gg{1}$, we introduce the HP creation and annihilation operators $\hat{b}^\dag_{j}$ and $\hat{b}_{j}$ ($j=1,..,M$) as
\begin{equation}
\label{HP}
\hat{S}^+_j=\sqrt{N}\hat{b}^\dag_{j}\hat{\cal A}_j,~~\hat{S}^-_j=\sqrt{N}\hat{\cal A}_j\hat{b}_{j},~~\hat{S}^z_j=\hat{b}^\dag_j\hat{b}_j-N/2,
\end{equation}
with $[b_{j},b^\dag_{l}]=\delta_{jl}$. Here, the operator $\hat{\cal A}_{j}=\sqrt{1-{\hat{b}^\dag_j\hat{b}_j}/{N}}$ is necessary in order to guarantee that the Dicke operators $\hat{S}^{\pm}_{j}$ and $\hat{S}^z_j$ satisfy the proper SU(2) algebra. In terms of effective HP bosons, Eq.~(\ref{hamiltonian}) takes the form
\begin{equation}
\hat{\cal H}_{HP}=-\sum^M_{j=1}{J}\hat{a}_{j}\hat{a}^\dag_{j+1}+\sum^M_{j=1}\Omega\sqrt{N}\hat{b}^\dag_{j}\hat{\cal A}_j\hat{a}_j+h.c.
\end{equation}
The enhancement of the intracavity coupling due to the collective nature of the  ensemble is clearly retrieved. The nonlinear nature of the interaction between the two bosons entering the dynamics is encompassed by the operators $\hat{\cal A}_{j}$, which depends on the number of excitations in the state of the effective HP particle. The strength of any nonlinear effect results from a trade-off between $\langle{b}^\dag_jb_j\rangle$ and $N$. For our discussion, it is useful to expand $\hat{\cal A}_{j}$ in power-series as 
%\begin{equation}
%\label{expansion}
$\hat{\cal A}_{j}=\sum^\infty_{l=0}\frac{(2l)!}{(1-2l)(2^ll!)^2}\frac{(\hat{b}^\dag_j\hat{b}_j)^l}{N^l}$.
%\end{equation}
Although any order of this expansion could be taken, for clarity of presentation we consider the situation of a ``mesoscopic'' number of atoms per ensemble, {\it i.e.} $N$ is such that $\hat{\cal A}_j\simeq{1}-{\hat{b}^\dag_j\hat{b}_j}/{2N}+{\cal O}(N^{-2}),\,\forall{j}$. Physically, this implies that the number of implanted two-level atoms per cavity has to be large enough for the HP transformation to be valid but sufficiently small not to blur any nonlinear effect. By introducing the truncated expressions for $\hat{\cal A}_j$'s into $\hat{\cal H}_{HP}$ we clearly identify the form the non-linear mechanism has, up to the considered order of approximation
\begin{equation}
\label{approx}
\begin{aligned}
\hat{\cal H}_{HP}&\simeq\sum^M_{j=1}(\Omega\sqrt{N}\hat{b}^\dag_j\hat{a}_j-J\hat{a}^\dag_j\hat{a}_{j+1}-\frac{\Omega}{2\sqrt{N}}\hat{b}^{\dag{2}}_j\hat{b}_j\hat{a}_j+h.c.)
\end{aligned}
\end{equation}
The structure of this Hamiltonian is interesting. The term $\Omega\sqrt{N}(\hat{b}^\dag_j\hat{a}_j+\hat{b}\hat{a}^\dag_j)$ 
%has the form of the generator of a beam-splitting operation between a cavity mode and the corresponding HP boson. Being 
is quadratic in the quadrature variables of the two modes and is obviously linear. On the other hand, the term $\Omega(\hat{b}^{\dag{2}}_j\hat{b}_j\hat{a}_j+\hat{b}^{\dag{}}_j\hat{b}^2_j\hat{a}^\dag_j)/2\sqrt{N}$ encompasses the nonlinear character of the interaction that, noticeably, is not of the cross-Kerr form~\cite{books}. It is worth stressing that, while in the usual approach to coupled-cavity problems, the hopping term in Eq.~(\ref{approx}) is diagonalized by means of a canonical transformation introducing collective cavity modes~\cite{generale}, here we keep its structure which is instrumental to the following discussion. We notice that, by assuming the conditions for adiabatic elimination of the atomic degrees of freedom, each $\hat{b}_j$ would effectively become proportional to $\hat{a}_j$ and Eq.~(\ref{approx}) would reduce to the Bose-Hubbard model~\cite{generale}. 
%Therefore, everything known on Mott transition can be adopted here

{\it Dynamical equations}.- It is straightforward to find the Heisenberg equations of motion for the two bosonic species involved in the dynamics. These read
\begin{equation}
\label{heisenberg}
\begin{aligned}
i\partial_t\hat{a}_{j}&=-J(\hat{a}_{j+1}+\hat{a}_{j-1})+\Omega\sqrt{N}\hat{b}_j-\frac{\Omega}{2\sqrt{N}}\hat{b}^\dag_{j}\hat{b}^2_j,\\                           i\partial_t\hat{b}_j&=\Omega\sqrt{N}\hat{a}_j-\frac{\Omega}{2\sqrt{N}}(2\hat{b}^\dag_j\hat{b}_j\hat{a}_j+\hat{b}^2_j\hat{a}^\dag_j).                            \end{aligned}
\end{equation}

The coupled nature and the discreteness of these nonlinear equations make their direct solution a formidable problem. In order to tackle such a challenge, we work in a way to reconduct Eqs.~(\ref{heisenberg}) into a manageable form, exploiting some plausible physical assumptions. First, we would like to eliminate the difficulties related to the non-commutativity of the operators in Eqs.~(\ref{heisenberg}). Now, if the operating temperature of the system is low, we can neglect the fluctuations of the operators $\hat{b}_j,\,\hat{b}^\dag_j$ calculated over the state the atoms are into and concentrate just on their mean values. Furthermore, we can take each cavity as prepared in a large-amplitude coherent states. This allows us to replace the operators appearing in the dynamical equations of motion with a corresponding complex scalar function~\cite{semiclass}. We therefore operate the substitutions $a_{l}\rightarrow\alpha_l,\,a^\dag_{l}\rightarrow\alpha^{*}_l$ and $b_{l}\rightarrow\beta_l,\,b^\dag_{l}\rightarrow\beta^{*}_l$, where each complex quantity is a function of time. Eqs.~(\ref{heisenberg}) now become
\begin{equation}
\label{semic}
\begin{aligned}
i\partial_t\alpha_{l}&=-J(\alpha_{l+1}+\alpha_{l-1})+\Omega\sqrt{N}\left(\beta_l-\frac{|\beta_l|^2\beta_l}{2N}\right),\\
i\partial_t\beta_l&=\Omega\sqrt{N}\alpha_l-\frac{\Omega}{2\sqrt{N}}(\beta^2_l\alpha^*_l+2\alpha_l|\beta_l|^2).
\end{aligned}
\end{equation}
As anticipated, these equations are easily generalizable to the case of arbitrary-order expansion in the nonlinear operator $\hat{\cal A}_j$. Our approach is designed in order to allow for such a generalization. So far, we retained the discreteness of the equations of motion. For a clear understanding of our results, it is more intuitive to proceed towards the {\it continuous limit} where the function $\gamma$ depends on a continuous position-variable $x$ along the linear array of cavities (here $\gamma=\alpha,\beta$ and their $h.c.$). This simply means that the inter-distance $d$ between two consecutive atom-cavity systems is much smaller than the scale represented by the wavelength $\lambda$ of the wave-like excitation propagating across the array. That is, by taking $k=2\pi/\lambda$, it must be $kd\ll{2\pi}$ so that $\gamma_j\rightarrow{\gamma}(x,t)$. In this way, the photon-hopping contribution to the first of Eqs.~(\ref{approx}) can be modified considering that $\alpha_{j+1}+\alpha_{j-1}\rightarrow2\alpha(x,t)+d^2\partial_{xx}\alpha(x,t)$. This changes the equation for $\alpha(x,t)$ into
%\begin{equation}
%\begin{aligned}
$i\partial_t\alpha(x,t)=-2J\alpha(x,t)-Jd^2\partial_{xx}\alpha(x,t)+\Omega\sqrt{N}[\beta(x,t)-{|\beta(x,t)|^2\beta(x,t)}/{2N}]$,
%i\partial_t\beta(x,t)&=\Omega\sqrt{N}\alpha(x,t)-\frac{\Omega}{2\sqrt{N}}(\beta^2(x,t)\alpha^*(x,t)\\
%&+2\alpha(x,t)|\beta(x,t)|^2),
%\end{aligned}
%\end{equation}
reminding of a Schr\"odinger equation with a cubic non-linearity (SCN)~\cite{agrawal}, which is known to admit solitonic solutions. However, the equation at hand is still quite different from an SCN as the nonlinear term couples the dynamics of the real bosons to that of the effective ones. For compactness, we drop any explicit dependence on the variables $x$ and $t$.

Rather than using brute force numerical methods for the solution of the problem (such as Newtonian relaxation techniques~\cite{meier}), we would like to get a clear picture of the physical process. We therefore decide to use an approach that is able to capture the essential features of the physics behind this problem. In Ref.~\cite{sipe}, the so-called {\it multiple-scale} technique has been used in order to investigate gap solitons in non-linear periodic structures. The flexibility of this technique and its successful application in~\cite{sipe} suggest its adaptation to the present situation. The method is based on the expansion of both time- and space-derivatives in terms of mutually independent time- and length-{scales} $v_p=\mu^pv$ ($v=t,\,x$) according to $\partial_{v}=\sum_{p\ge{0}}\mu^p\partial_{v_p}$. Similarly, $\gamma=\sum_{p\ge{1}}\mu^p\gamma^{(p)}$,
% ($\gamma=\alpha,\beta,\alpha^*,\beta^*$). 
where we have introduced the small parameter $\mu$. In its essence, the multiple-scale method utilizes a separation of spatial and temporal scales in analogous to other standard techniques used in quantum optics such as the slowly varying envelope approximation~\cite{books}.
% and the elimination of secular terms. 
Following~\cite{sipe}, we stop the expansion at order $p=3$ in $\gamma$, which is enough, in the conditions of small nonlinearity at hand, to encompass the important aspects of the system's dynamics. We now proceed to solve the problem corresponding to a scale of order $p-1$ and use it into the one at order $p$. 
%The method is useful because it allows for the easy inclusion of any order to approximation in ${\cal A}_j$. 
By replacing the scale-expansion into the equations of motion and collecting terms corresponding to the same power of $\mu$, we get the {\it universal structures}
%\begin{equation}
%\begin{aligned}
$(i\partial_{t_0}+2J+Jd^2\partial_{x_0x_{0}})\alpha^{(p)}-\Omega\sqrt{N}\beta^{(p)}=\Gamma^{(p)}_{\alpha},\,i\partial_{t_0}\beta^{(p)}-\Omega\sqrt{N}\alpha^{(p)}=\Gamma^{(p)}_{\beta}$
%\end{aligned}
%\end{equation}
%where the different $x_{p}$ and $t_p$ variables have been treated as mutually independent and
with $\Gamma^{(p)}_{\alpha,\beta}$ depending on higher-order derivatives of $\alpha^{(p-k)}$ and $\beta^{(p-k)}$ ($k\in{\mathbb Z}$). The study is performed stepwise: At order $\mu$ we have $\Gamma^{(1)}_{\alpha,\beta}=0$ so that $\alpha^{(1)}$ and $\beta^{(1)}$ satisfy linear equations in the {\it slow} variables $x_0$ and $t_0$ which are combined together to give
\begin{equation}
%\begin{aligned}
\label{wave}
(\partial_{t_0t_0}-2iJ\partial_{t_0}-iJd^2\partial_{t_0x_0x_0}+\Omega^2{N})\alpha^{(1)}=0
%\end{aligned}
\end{equation}
with $\beta^{(1)}=-i\Omega\sqrt{N}\int\alpha^{(1)}{d}t_0$. In order to solve Eq.~(\ref{wave}) we take the plane-wave ansatz $\alpha^{(1)}={\cal E}{\rm e}^{i(kx_0-\omega{t}_0)}$ with ${\cal E}$ an envelope function depending on space and time scales faster than $x_0$ and $t_0$ and $\omega$ the frequency of such a carrier excitation. The associated solvability condition leads to the dispersion relation
%\begin{equation}
%\label{lwdisp}
$\omega_{\pm}(k)=-J(1-{d^2k^2}/{2})\pm\sqrt{J^2(1-d^2k^2/2)^2+\Omega^2N}$,
%\end{equation}
thus defining effective ``optical'' and ``acoustic'' branches (corresponding to $\omega_+(k)$ and $\omega_-(k)$ respectively) reminding us of a linear diatomic crystal. The difference with our case is that the two species in the crystal ({\it i.e.} true and effective bosons) share the same site-location. We remark that $\omega_\pm(k)$ are the long-wavelength approximations of the branches coming from the exact dispersion relation for the {\it discrete} equations of motion (\ref{heisenberg}) in the linear limit. This reads
%\begin{equation}
%\label{disp}
$\tilde\omega_{\pm}(k)=-J\cos(kd)\pm\frac{}{}\sqrt{J^2\cos^2(kd)+\Omega^2N}$,
%\end{equation}
which becomes $\omega_{\pm}(k)$ for $kd\ll{1}$ and exhibits a band-gap at the edge of the first Brillouin-zone~\cite{discrete}. Here, in order to be clear, we concentrate on the optical branch.

The previous level of solution leaves ${\cal E}$ unknown. Its form is determined by going to higher order in $p$ and imposing the appropriate solvability conditions (such as the nullity of any secular term, as required in order for the multiple-scale approach to hold~\cite{sipe}). At order $\mu^{2}$ this results in the equation $(\partial_{t_1}+v_{g_+}\partial_{x_1}){\cal E}=0$, so that the envelope function must depend on the variable $\xi=x_1-v_{g_+}t_1$ with the group velocity $v_{g_+}=\frac{2kJd^2\omega^2_{+}}{(\omega^2_++\Omega^2N)}\simeq\frac{d\omega_+}{dk}$. 
%(to leading order in $kd$). 
The iteration of such an approach finally leads to the generalized SCN
\begin{equation}
\label{SCN}
i\partial_{t}\epsilon+c_1\partial_{\chi\chi}\epsilon+c_2|\epsilon|^2\epsilon=0.
%i\partial_{t_2}{\cal E}+c_1\partial_{\xi_\pm\xi_\pm}{\cal E}+c_2|{\cal E}|^2{\cal E}=0,
\end{equation} 
where $c_1=(Jd^2\omega^3_++\Omega^2Nv^2_{g_+})/[{\omega_+(\omega^2_++\Omega^2N)}]$, $c_2={2\Omega^4N}/[{\omega_+(\omega^2_++\Omega^2N)}]$, ${\cal E}=(1/\mu){\epsilon}$, $\chi=\xi/\mu$ and $t_2=\mu^2{t}$, as implied by the multiple-scale definition. Eq.~(\ref{SCN}) can be exactly solved by means of well-known inverse-scattering methods (ISM) and is known to admit the solitonic solution 
%$\epsilon=\sqrt{2c_1/c_2}\eta~{\rm sech}\{\eta[x-(v_{g+}+2c_1\sigma)t]-\nu\}{\rm e}^{i\sigma(x-v_{g+}t)-ic_1(\sigma^2-\eta^2)t-i\phi_0}$ 
$\epsilon=\eta\sqrt{2c_1/c_2}{\rm sech}\{\eta[\chi-2c_1\sigma{t}]-\nu\}{\rm e}^{i\sigma(\chi-ic_1(\sigma^2-\eta^2)t-i\phi_0}$ 
for ${c_1c_2}>0$~\cite{SCN}, which is the case here. {\it We have thus retrieved a solitary-wave behavior in the linear array being studied which was the central aim of our study}. Here, $\eta,\,\sigma,\,\nu$ and $\phi_0$ are integration constants to be determined from the boundary conditions attributed to each specific physical problem. The inclusion of higher-order terms in the expansion of $\hat{\cal A}_j$'s (implying the use of lesser atoms per ensemble) results in the modification of $\Gamma^{(p)}_{\alpha,\beta}$ and, consequently, of $c_{1,2}$. Multi-soliton solutions can also be found through the ISM theory~\cite{SCN}.
%\begin{equation}
%\end{equation}
\begin{figure}[b]
\centerline{{\bf (a)}\hskip3.5cm{\bf (b)}}
\centerline{\includegraphics[width=0.47\textwidth]{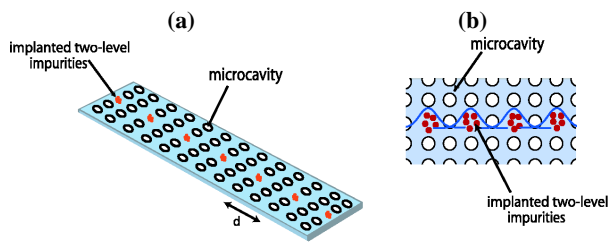}}
\caption{Sketch of the physical setup. {\bf (a)}: A photonic crystal with defect microcavities (formed every other site). The microcavity spacing is $d$. {\bf (b)}: Each cavity contains $N$ impurities and is coupled to its nearest-neighbors via photon-hopping mechanism.}
%due to the overlap of the evanescent part of the cavity field.}
\label{fig:fig1}
\end{figure}
The transition from solitonic dynamics to the linear regime 
%(where ${\cal E}$ is just a constant or a wave-packet, as determined by the dispersion relation) 
occurs for $c_{1}\gg{c_2}$. In the long-wavelength approximation and in the realistic situation of $J\ll\Omega$, this corresponds to $\sqrt{N}\gg\Omega/J$, which imposes limitations on the number of atoms per cavity, in agreement with our analysis above. As an important remark, we notice that the continuous limit adopted here for clarity of exposition can be bypassed by approaching the discrete dynamical equations so as to find a solution valid within the whole Brillouin zone. This can be done with minimal changes to our approach and without affecting the conclusions of our study by taking $\gamma_l=\sum_{p}\mu^p\gamma^{(p)}(\xi_l,\tau,\phi_l)$ with $\xi_{l}=\mu(ld-\tilde\lambda{t})$, $\tau=\mu{t}$, $\phi_{l}$ representing a phase difference between each system and $\tilde\lambda$ to be found via the solvability conditions~\cite{notasolvability}. 

{\it Physical setup}.- We now briefly discuss a potential candidate for the observation of the predicted phenomena, which we identify in an array of coupled photonic crystal microcavities~\cite{pc}. A sketch is provided in Fig.~\ref{fig:fig1}. We stress that the linear arrangement addressed here can be generalized with some efforts to a bidimensional configuration by replacing the cavity adjacency-matrix, which determines the hopping term in the first of Eq.~(\ref{heisenberg}), with an appropriate tensor. Each cavity is formed as a single defect (created, for instance, every other site) in the lattice-pattern of the crystal. The spacing between the cavities can be made as small as tenth of the wavelength of the radiation confined in the crystal (which needs to be a slab of thickness $\sim0.5d$ in order to confine radiation on the plane~\cite{pc}). The microcavities are doped with two-level impurities (for instance, substitutional Si donor impurities in a GaAs photonic crystal), $N$ being determined by properly adjusting the bulk doping density, $N\simeq{10}$ being experimentally realistic~\cite{yama}. As stated, the small mode-volume of each cavity allows for large atom-field coupling rates. A proper pattern of the array, together with improved quality of the resonators, allow for $\Omega/J\sim10$. The effects of atomic spontaneous emission and 
%the cavities open dynamics (due to their 
relatively small cavity quality-factor in the formation of solitary waves is beyond the scopes of this work and is the focus of ongoing investigation.

{\it Remarks}.- We have addressed the dynamics of an array of coupled cavities from the perspective of intrinsic nonlinear optics. The nonlinearity arises from the coupling of the cavities with ensembles of two-level atoms. Such built-in nonlinearity and the periodic photon tunneling compete so as to create solitary waves. We have used a simple iterative method to demonstrate such a behavior in a system of coupled photonic crystal microcavities. Our study contributes to the affirmation of systems of coupled cavities as flexible quantum simulators and represents an original approach to their dynamics, interestingly counterposed to current mainstream~\cite{generale}.

{\it Acknowledgments}.- MP thanks G.M. Palma, C. Di Franco and M.S. Tame. We acknowledge support from The Leverhulme Trust (ECF/40157), UK EPSRC and QIPIRC. GSA is supported by NSF (Grant~No.~CCF0524673).


\begin{thebibliography}{99}
\bibitem{notasolvability} Somehow obviously, $\tilde\lambda$ is the group velocity of the excitation.
\end{thebibliography}
\end{document}